\documentclass[peerreview]{IEEEtran} 
\usepackage[ansinew]{inputenc}
\usepackage{epsfig}
\usepackage{amsfonts}
\usepackage{amssymb}
\usepackage{amsmath}
\usepackage{subfigure}
\usepackage{color}
\usepackage{comment}
\usepackage{multirow}
\usepackage{setspace}

\begin{document}

\title{On the Performance of Packet Aggregation \\in IEEE 802.11ac MU-MIMO WLANs}
\author{Boris Bellalta,~\IEEEmembership{Member,~IEEE,} Jaume Barcelo, Dirk Staehle, Alexey Vinel,~\IEEEmembership{Senior Member,~IEEE,} Miquel Oliver,~\IEEEmembership{Member,~IEEE}
\thanks{Boris Bellalta, Miquel Oliver are with Universitat Pompeu Fabra, Spain (email:\{boris.bellalta,miquel.oliver\}@upf.edu), Jaume Barcelo is with the Universidad Carlos III Madrid, Spain (email:jaume.barcelo@uc3m.es), Dirk Staehle is with the University of Wurzburg, Germany (email:\{dstaehle\}@informatik.uni-wuerzburg.de) and Alexey Vinel is with Tampere University of Technology, Finland (alexey.vinel@tut.fi). \textcolor{blue}{This paper has been accepted for publication in IEEE Communication Letters. July 2012.}}}
    
\date{}

\maketitle

\begin{abstract}
Multi-user spatial multiplexing combined with packet aggregation can significantly increase the performance of Wireless Local Area Networks (WLANs). In this letter, we present and evaluate a simple technique to perform packet aggregation in IEEE 802.11ac MU-MIMO (Multi-user Multiple Input Multiple Output) WLANs. Results show that in non-saturation conditions both the number of active stations (STAs) and the queue size have a significant impact on the system performance. If the number of STAs is excessively high, the heterogeneity of destinations in the packets contained in the queue makes it difficult to take full advantage of packet aggregation. This effect can be alleviated by increasing the queue size, which increases the chances of scheduling a large number of packets at each transmission, hence improving the system throughput at the cost of a higher delay.
\end{abstract}



\doublespacing

\section{Introduction} \label{Sec:Intro}

The upcoming IEEE 802.11ac standard for WLANs promises throughputs higher than $1$ Gbps in the $5$ GHz band. Compared with the IEEE 802.11n standard \cite{80211n}, IEEE 802.11ac considers wider channels, modulation and coding rates with higher spectral efficiency and MU-MIMO capabilities, as well as channel bonding mechanisms \cite{ong2011ieee,park2011ieee}. The IEEE 802.11ac Multiple Access Control (MAC) layer will basically follow the IEEE 802.11n standard, only extending it to accommodate the new MU-MIMO features, i.e., the ability to transmit multiple spatial streams from the Access Point (AP) to different STAs in parallel by using a multi-user beamforming scheme.
  
One of the key features of the IEEE 802.11n MAC layer is the ability to aggregate packets in order to reduce temporal overheads (interframe spaces, MAC and Physical (PHY) layer headers and preambles) that significantly harm the performance of WLANs \cite{li2009aggregation,ong2011ieee}. The same benefits of packet aggregation are expected in MU-MIMO enhanced WLANs, although specific considerations to implement it have to be addressed. Mainly, if multi-user beamforming is applied at the AP, each STA receives only the  spatial streams directed to it and therefore, only packets directed to the same STA can be aggregated in those spatial streams. 

In this letter, we propose and evaluate a simple reference scheme covering the fundamental properties of packet aggregation and MU-MIMO transmission in order to demonstrate that the combination of both techniques is able to significantly improve the system performance. In particular, the contributions of this letter are: $1$) present a new mechanism based on the RTS/CTS handshake to both signal the selected STAs and perform explicit channel sounding, $2$) quantify the gain in performance that packet aggregation in IEEE 802.11ac  WLANs can provide and highlight the effect of the buffer size on the attainable throughput, and $3$) provide guidelines for buffer dimensioning to maximize the system performance.


\section{A Joint Spatial Multiplexing and Packet Aggregation scheme} \label{Sec:Model}

\subsection{System Model}

A single access point (AP) equipped with $M$ antennas, each one with its corresponding radio-frequency (RF) chain, and a finite buffer space of $K$ packets is considered. Packets with a constant length of $L_d$ bits and destined to $N$ single-antenna STAs arrive at the AP following a Poisson process of rate $\lambda$ packets/second. The probability that the AP receives a packet directed to a certain STA is the same for all STAs and equal to $1/N$. All active destinations share a single finite-buffer space. If traffic differentiation issues are not considered, the use of a single shared finite-buffer results in a higher system performance compared to the use of multiple independent queues of size $K/N$ \cite{Kleinrock_CSCP}.

By using a multi-user beamforming scheme, the AP is able to create $m\in [1,M]$ spatial streams at each transmission, each one directed to a different STA. It is assumed that the channel does not introduce channel errors and that the same transmission rate can be used to send data to all STAs. Explicit Channel State Information (CSI) \cite{gong2010training} from each STA is obtained at each transmission by extending the RTS/CTS frames. We use the notation RTS$^*$ and CTS$^*$ to refer to such extended signaling frames. In case that the CSI for the selected STAs is not outdated, such procedure could be avoided, thus reducing the required overheads. However, in this letter, we have assumed that CSI estimation and reporting is performed at each transmission. In addition, it is assumed that only the AP is transmitting and the $N$ active users act only as receivers. 

\begin{figure*}[tt!!!!!!!!]
\centering
\epsfig{file=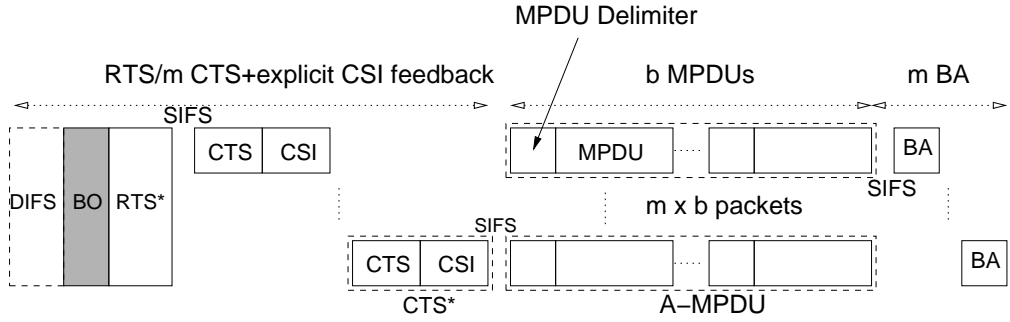,scale=0.6,angle=0}
\caption{Transmission of $b\cdot m$ packets through $m$ spatial streams following the MAC protocol considered in this letter.  RTS$^*$ and CTS$^*$ are the extended versions of the RTS and CTS frames respectively.}
\label{Fig:Space_batch}
\end{figure*}

\subsection{Channel Access}

It is expected that the IEEE 802.11ac medium access control (MAC) will be similar to that of the IEEE 802.11n, including the explicit CSI feedback as a main mechanism for channel sounding. The use of the RTS/CTS mechanism is considered to: $1$) signal the selected STAs at each transmission, $2$) protect a large A-MPDU transmission from collisions and 3) perform the channel estimation and explicit CSI reporting by the STAs. In detail, the proposed access scheme for packet aggregation is described as follows and depicted in Figure \ref{Fig:Space_batch}. After an initial DIFS and backoff (BO) periods, the AP sends a RTS$^*$ that contains the address of those STAs that have been selected as potential receivers of an A-MPDU frame. Using the $M$ Very-high Throughput Long Training Fields (VHT-LTF) included in the PHY preamble of the RTS$^*$, STAs are able to estimate the required CSI information (fading coefficients for each data subcarrier and for each antenna). Then, the selected STAs reply to the AP with an extended CTS$^*$ to confirm that they are ready and feed the AP with fresh CSI. The replies of the STAs follow the same order as the addresses in the RTS$^*$ packet. In each spatial stream, an A-MPDU frame composed by $b$ MPDU packets is transmitted. After the reception of the different spatial streams, each STA replies with a Block ACK (BA) that contains a bit pattern field to indicate the position of the correctly received packets. Packets that are not positively acknowledged remain in the queue until they are successfully transmitted. 
 
\subsection{Packet Aggregation Reference Scheme}

At each transmission, $l$ packets are sent to the medium using $m\in[1,M]$ spatial streams and including $b=l/m,~b\in[1,B]$ packets per stream, with $B$ the maximum A-MPDU size. Given that there are $q$ packets waiting for transmission at the AP, the packets included in the next transmission are selected as follows: First, the  number of spatial streams that will be scheduled is fixed to $m = min(\xi,M)$, where $\xi$ is the number of different destinations among the packets stored at the AP. Then, the number of packets per spatial stream is set to $b=min(\psi,B)$, where $\psi$ is the highest number of packets such that $m$ different destinations have at least $\psi$ waiting packets at the AP. This is equivalent to sorting all the STAs with pending packets for transmission in descending order, starting from the one with more packets, and then set $b$ as the number of waiting packets destined to the $m$-th STA. In the case that more than $M$ different destinations have at least $\psi$ waiting packets, the $M$ selected STAs are chosen following a FIFO policy with respect to the first waiting packet. 

As a single transmission rate is considered in this paper, the requirement that all the spatial streams include the same number of packets implies that the payload in all of them has the same duration, allowing to maximize the parallelization capabilities of MU-MIMO. With multiple transmission rates, the packet aggregation mechanism can benefit from considering multiuser diversity \cite{viswanath2002opportunistic} to select the STAs with favourable channel conditions at each transmission.

\subsection{Fairness}

The packet aggregation scheme proposed in this paper is designed for throughput maximization. It tries to send as many packets as possible at each transmission, prioritizing the STAs with more packets waiting in the queue. In such situation, the scheme proposed is fair if the traffic load is evenly distributed among all the STAs, as it gives the same transmission opportunities to all of them. However, if the traffic load is not balanced, the STAs receiving less traffic may suffer from higher delays as transmissions directed to them are scheduled less often. Nevertheless, it is straightforward to modify the scheduling mechanism to provide fairness at the expense of throughput. One option would be to always serve the STA to which the first packet of the queue is directed. Another alternative is to time-stamp packets and take into account packet aging for scheduling decisions.

\subsection{Transmission delay for an A-MPDU frame}

Each transmission has an average duration of $T(m,b)=T_{BO}+\text{DIFS}+T_{\text{RTS}^*}+ m(\text{SIFS}+T_{\text{CTS}^*})+T_{A}(b)+m(\text{SIFS}+T_{\text{BA}})$ seconds, and depends on the number of spatial streams $m$ and the number of packets aggregated, $b$, in each spatial stream. In detail, 

\begin{eqnarray}
	T_{\text{RTS}^*}=&P_{\text{vht}}(M)+\left \lceil \frac{\text{SF} + L_{\text{RTS}^*} + \text{TB}}{L_{\text{DBPS}}} \right \rceil T_s\\
	T_{\text{CTS}^*}=&P_{\text{vht}}(1)+\left \lceil \frac{\text{SF} + L_{\text{CTS}^*} + \text{TB}}{L_{\text{DBPS}}} \right \rceil T_s\\
	T_A(b)=&P_{\text{vht}}(M)+\left \lceil \frac{\text{SF} + b (\text{MD}+\text{MH}+L_d) + \text{TB}}{L_{\text{DBPS}}} \right \rceil T_s\\
	T_{\text{BA}}=&P_{\text{vht}}(1)+\left \lceil \frac{\text{SF} + L_{\text{BA}} + \text{TB}}{L_{\text{DBPS}}} \right \rceil T_s
\end{eqnarray}
where $T_{BO}$ is the average backoff duration, $P_{\text{vht}}(w)=36+T_s w~\mu$s is the duration of the PHY-layer preamble and headers, with $w$ the number of VHT-LTF fields included in it and used for channel estimation (i.e, $w=M$) and $T_s=4~\mu$s the duration of an OFDM symbol. $\text{SF}$ is the \textit{service field} ($16$ bits), $\text{TB}$ are the \textit{tail bits} ($6$ bits), $L_{\text{DBPS}}$ is the number of bits in each OFDM symbol, $\text{MD}$ is the \textit{MPDU Delimiter} ($32$ bits, only used if $b>1$) and $\text{MH}$ is the \textit{MAC header} ($288$ bits). The length of the RTS$^*$, CTS$^*$ and BA frames is  $L_{\text{RTS}^*}=160+46(M-1)$ bits, $L_{\text{CTS}^*}=112+L_{\text{CSI}}$ bits and $L_{\text{BA}}=256$ bits respectively. We have used the values of \cite{ong2011ieee} when possible. For those values not defined in \cite{ong2011ieee}, we have taken the values from the IEEE 802.11n standard \cite{80211n}.

\subsection{Example}

\begin{figure*}[tt!!!!!!!!]
\centering
\epsfig{file=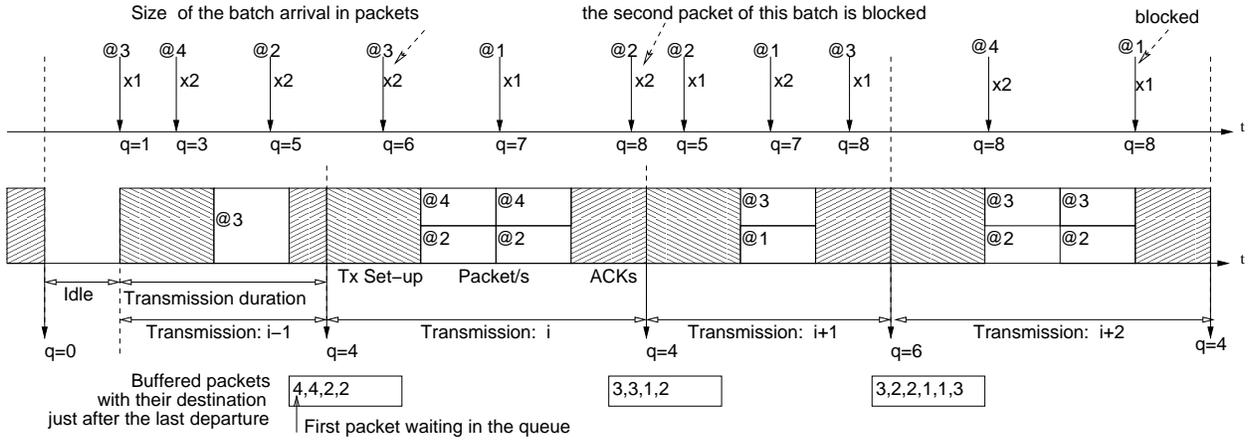,scale=0.50,angle=0}
\caption{Temporal evolution of the AP's queue with $M=2$ antennas, $B=2$ and $K=8$. The $q$ value in the figure represents the queue state (number of frames in the queue) just after a batch arrival (i.e. several packets that arrive together to the AP's queue) or a transmission departure. The $@x$ refer to the destination of the packets in a batch.}
\label{Fig:Temporal}
\end{figure*}

In Figure \ref{Fig:Temporal}, a specific example of the system operation is shown for $M=2$ antennas, $B=2$ and a finite buffer space of size $K=8$ packets shared among all active STAs (a small buffer size has been considered in this example to illustrate that packets are lost when they arrive at a full buffer). The $(i-1)$-th transmission comprises a single packet as it is scheduled as soon as a new packet arrives to the AP. During the $(i-1)$-th transmission, two packets directed to the fourth and two packets directed to the second STAs are buffered and assembled together in the $i$-th transmission after the end of the $(i-1)$-th one. Similarly, during the $i$-th transmission two packets directed to the third STA arrive at the AP, a single packet directed to the first STA, as well as two directed to the second STA, one of which is blocked because there is no free space in the buffer. Observe that, when the $(i+1)$-th transmission is scheduled, there are four frames in the transmission buffer, two directed to the third STA, one directed to the first STA and one directed to the second STA. Then, following the previous described algorithm, as $\psi=1$ only two packets are transmitted, each one using a different spatial stream. Finally, when the $(i+2)$-th transmission is scheduled, the AP has two packets directed to the third STA, two packets directed to the second STA and two packets directed to the first STA. In such situation, the AP selects the packets directed to the third and second STAs by applying the FIFO policy with respect to the first waiting packet, which in this case correspond to one packet directed to the third and one to the second destination.


\section{Performance Evaluation} \label{Sec:Results}

\begin{table}[t!!]
\centering

 \begin{tabular}{|c|c|c|}
   \hline
  Parameter & Notation & Value \\
  \hline
  Number of bits / OFDM symbol & $L_{\text{DBPS}}$ & $1560$ bits \\
  CSI feedback & $L_{\text{CSI}}$ &  $1872 M$ bits \\
  Packet Length & $L_d$ & $12000$ bits \\
  Max. A-MPDU size & B & $64$ packets \\
  Av. Backoff Duration & $T_{BO}$ & $139.5\mu$\\
  \hline
 \end{tabular}
 \caption{Parameters considered for the evaluation of the proposed system \cite{80211n,ong2011ieee}.}\label{Tbl:parameters}
\end{table}

In this section, the performance of our scheme is evaluated in terms of the amount of packet losses (blocking probability) and delay for different traffic loads, different number of active STAs and several number of antennas and buffer sizes at the AP. Simulation results are compared against the queueing model presented and validated in \cite{BellaltaMSWIM2009}, which gives the upper-bound performance for the proposed packet aggregation scheme in MU-MIMO systems when Poisson arrivals and a finite buffer is considered at the transmitter. The simulator has been built from scratch using C++ and it is based on the COST (Component Oriented Simulation Toolkit) libraries \cite{CostSim}. The specific parameter values considered are given in Table \ref{Tbl:parameters}. The number of bits in each OFDM symbol has been computed assuming a $80$ MHz channel bandwidth, a QAM-$256$ modulation and a $5/6$ coding rate. The required bits for the CSI feedback is computed assuming that $16$ bits are required for each $2$ OFDM data subcarriers and transmitting antenna. The average duration of the backoff $T_{BO}$ has been computed considering an average backoff value of $15.5$ slots and a slot duration of $9\mu$s.

\subsection{Maximum Performance}

The maximum achievable throughput by the AP is obtained when $M$ A-MPDU of $B$ packets each one are continuously transmitted, and is given by 

\begin{equation}
	S_{max}(M,B)=\frac{M\cdot B\cdot L_d}{T(M,B)}
\end{equation}

For instance, with $M=4$ antennas and no packet aggregation ($B=1$ packets), the maximum system throughput is equal to $S_{max}(4,1)=55$ Mbps, which increases to $S_{max}(4,64)=1070$ Mbps when packet aggregation is enabled ($B=64$ packets).

The previous expression does not consider the influence of the buffer size nor the traffic arrival process at the AP. This is solved by using, properly parametrized, the queueing model presented in \cite{BellaltaMSWIM2009}. It gives the maximum throughput, minimum blocking probability and minimum average delay in a MU-MIMO system for the proposed packet aggregation scheme. To obtain such performance metrics, the queueing model assumes that given $q$ packets waiting in the transmission queue, always $m=min(q,M)$ spatial streams and $b=\min\left(\left\lfloor \frac{q}{m} \right\rfloor, B\right)$ packets in each one can be aggregated, thus maximizing the number of packets that are transmitted at each attempt. 

\subsection{Results}

\begin{figure}[ttttt!!!!!!!!]
\centering
\subfigure[Blocking Probability.]{\epsfig{file=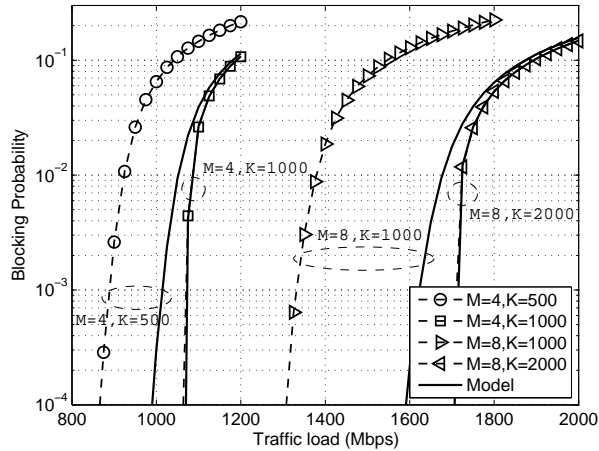,scale=0.55,angle=0}\label{Fig:BP_TL}}\\
\subfigure[Average number of Spatial Streams and average A-MPDU size ($M=4$ antennas). The x-axis is in log-scale.]{\epsfig{file=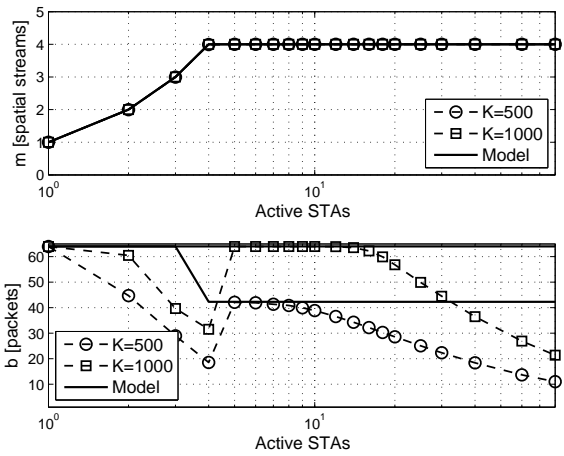,scale=0.55,angle=0}\label{Fig:EB_U}}
\subfigure[Delay ($M=4$ antennas). The x-axis is in log-scale.]{\epsfig{file=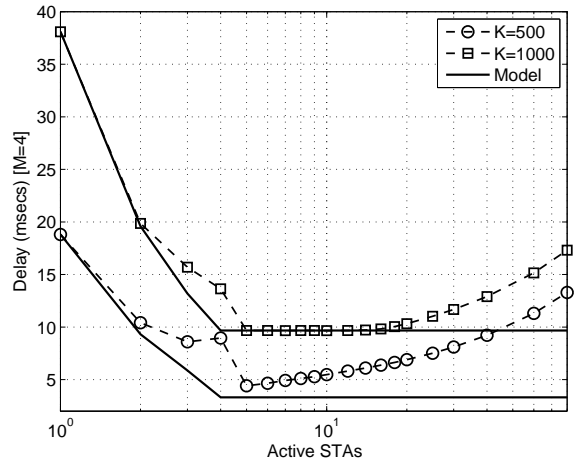,scale=0.55,angle=0}\label{Fig:ED_U}}
\caption{Performance of the packet aggregation scheme with different number of active STAs}
\end{figure}

In Figure \ref{Fig:BP_TL} the blocking probability (i.e. probability that a packet is discarded at its arrival at the AP) is plotted against the aggregate traffic load for $M=4$ and $8$ antennas. In both cases the number of active users is twice the number of antennas ($N=2M$). Increasing the number of antennas at the AP results in higher overheads, mainly related to longer RTS/CTS and acknowledgement phases. However, as the number of packets that can be sent at each transmission is also higher, the overall system performance is significantly improved, particularly if the buffer size is also properly increased. For $K=1000$ and a blocking probability equal to $10^{-2}$, doubling the number of antennas while keeping the same buffer size results in an increase of the supported load from 1098 to 1390 Mbps. In the case that the buffer size is also doubled, the supported load increases up to $1740$ Mbps. 

In Figure \ref{Fig:EB_U}, the average number of spatial streams allocated at each transmission and the number of packets aggregated in each spatial stream (A-MPDU size) are shown for different numbers of active STAs. In the same conditions, the average system delay is shown in Figure \ref{Fig:ED_U}. In both Figures, the AP has been equipped with $M=4$ antennas and the aggregate load has been fixed to $930$ Mbps and $1098$ for $K=500$ and $1000$ packets respectively. Those values are the traffic loads for which the blocking probability in Figure \ref{Fig:BP_TL} is equal to $10^{-2}$. The point at which the total number of aggregated packets in a single transmission ($l=m\cdot b$) is maximum is the point where the delay shows its minimum value. To operate at this optimal point, the number of active STAs has to be high enough to guarantee that always the maximum number of spatial streams is allocated and, at the same time, low enough to guarantee that there will be enough packets directed to the same STA in order to aggregate as many packets as possible. The counterpart of increasing the buffer size is an increase of the average system delay.

Previous results show the importance of properly sizing the buffer to achieve the maximum performance that the proposed joint spatial multiplexing and packet aggregation scheme can provide. Intuitively, the minimum buffer size has to be larger than $M\cdot B$ packets as, otherwise, the number of packets that can be aggregated is limited by the queue size. However, given that the packets arrive randomly at the AP, the buffer size must be several times larger to ensure that there are at least $M$ STAs that have at least $B$ packets each in the queue. The buffer size can be formulated in terms of $K=\alpha(M\cdot B)$, with $\alpha$ a design parameter that depends on the number of STAs, number of antennas, the packet aggregation parameters and the traffic load. In our particular case, with $M=4$ antennas, we have shown that with $\alpha=4$ the system performance is close to the optimal in the range between $5$ and $12$ users. In those cases in which the buffer is large enough to allow the scheduling of the maximum number of packets at each transmission, the blocking probability obtained from the simulation is close to the one derived from the queueing model. On the contrary, the difference between the simulation results and the queueing model show the inefficiencies in which a joint spatial multiplexing and packet aggregation scheme can incur when a smaller than required buffer is considered. 

%
%

\section{Conclusions} \label{Sec:Conclusions}

A basic scheme to perform packet aggregation in MU-MIMO IEEE 802.11ac WLANs has been presented and evaluated in non-saturation conditions. Results have shown that using packet aggregation the system performance is significantly increased, specially when the number of active STAs is only slightly higher than the number of antennas, and the buffer size is large enough to cope with the required heterogeneity of packet destinations. Under those conditions, the proposed scheme is close to optimal in terms of throughput as it is able to maximize both the number of spatial streams and the number of packets included in an A-MPDU frame.

\section*{Acknowledgment}

The authors want to express their gratitude to the anonymous reviewers and the editor for their helpful comments. This work has been partially supported by the Spanish Government under projects TEC2008-06055 (Plan Nacional I+D), CSD2008-00010 (Consolider-Ingenio Program), the Catalan Government (SGR2009\#00617) and WINEMO IC0906 COST Action.

%

\bibliographystyle{unsrt}

\end{document}